\documentstyle[aps,12pt]{revtex}
\begin{document}
\title{Scale-invariance of soil moisture variability and its implications for 
the frequency-size
distribution of landslides}
\author{Jon D. Pelletier, Bruce D. Malamud, Troy Blodgett, and Donald L. Turcotte}
\address{
  Department of Geological Sciences, Snee Hall, Cornell University \\
  Ithaca, NY 14853 \\
}
\maketitle
\newpage
\begin{abstract}
Power spectral analyses of soil moisture variability 
are carried out from scales of 100 m to 10 km on the microwave remotely-sensed
data from the Washita experimental watershed during 1992. 
The power spectrum $S(k)$ has an approximately power-law dependence
on wave number $k$ with exponent $-1.8$.   
This behavior is consistent with the behavior of a stochastic
differential equation for soil moisture at a point.
This behavior has important consequences for the frequency-size 
distribution of landslides. 
We present the cumulative 
frequency-size distributions of landslides induced by 
precipitation in Japan and Bolivia as well as landslides triggered
by the 1994 Northridge, California earthquake.
Large landslides in these regions, despite being triggered by
different mechanisms, have a cumulative
frequency-size
distribution with a power-law dependence on area with an
exponent ranging from $-1.5$ to $-2$.
We use a soil moisture field with the above statistics in conjunction
with a slope stability analysis to model the frequency-size distribution
of landslides.  
In our model
landslides occur when a threshold shear stress dependent on cohesion,
pore pressure, internal friction and slope angle is exceeded. 
This implies a threshold dependence on soil moisture and slope angle
since cohesion,
pore pressure, and internal friction are primarily dependent on soil
moisture. 
The cumulative
frequency-size distribution of domains of shear stress greater than a threshold
value with
soil moisture modeled as above and topography modeled as a Brownian walk
is a power-law function of area with
an exponent of $-1.8$ for large landslide areas.
This distribution is similar
to that observed for landslides. 
The effect of strong ground motion from earthquakes lowers the
shear stress necessary for failure but does not change the
frequency-size distribution of failed areas. This is consistent
with observations. 
This work suggests that remote sensing of soil moisture can be
of great importance in monitoring landslide hazards and proposes a
specific quantitative model for landslide hazard assessment.
\end{abstract}

\section{Introduction}
Landslides can be triggered by intense rainfall, rapid snowmelt, a change
in water level, and ground motion from earthquakes and volcanic eruptions
(Wieczoreck, 1996). The first three mechanisms act to increase 
soil moisture. The coincidences of increased groundwater levels and pore
pressures with landslides 
have been investigated by many authors (Fukuoka, 1980).
Studies suggest that a threshold dependence on soil moisture is appropriate
for slope instablity since landslides often occur when the product of rainfall
duration and intensity exceeds a threshold value (Caine, 1980; Wieczoreck, 1987).
{\it In situ} monitoring of pore pressure has shown that increases in pore
pressure resulting from heavy precipitation is coincident with landslides
(Johnson and Sitar, 1990). Several authors have shown that 
variations in the frequency of occurrence of landslides respond to climatic
changes with higher rates of landslides associated with wetter climates
(Grove, 1972; Innes, 1983;
Pitts, 1983; Brooks and Richards, 1994). Further evidence for 
correlations between landslides and
soil moisture is the observation that soil drainage
is the most successful method of landslide prevention in the United States
(Committe in Ground Failure Hazards, 1985).

Seismic triggering of landslides has also been studied by many authors
(Keefer, 1984; Jibson, 1996). 
Correlations between landslides and earthquakes can be inferred in a number
of ways. Many earthquakes have been known to trigger large numbers of
landslides and landslide densities have a strong correlation with 
active seismic belts 
(Jibson and Keefer, 1988; Tibaldi et al., 1995).
In some cases landslides have been associated with both high soil moisture
and earthquakes
(Seed, 1968). 

Landslides are a serious natural hazard and also 
play an important role in the
long-term shaping of landforms (Beck, 1968; Keefer, 1994; Densmore et al., 1997). In
this paper we investigate the frequency-size distribution of landslides
in areas where different triggering mechanisms dominate. We argue that
fractal models are useful for modeling the soil moisture and topographic
variations that are associated with the landslide instability.
A number of authors have examined the frequency-size distribution of 
landslides (Whitehouse and Griffiths, 1983; Ohmori and Hirano, 1988; Sugai, 1990; 
Sasaki, 1991; Hovius et al., 1997). 
These authors have consistently noted that the distribution 
is power-law or fractal. This is precisely analagous to the Gutenberg-Richter
law for earthquakes which states that the cumulative frequency of events with
a seismic moment greater than $M_{o}$ is a power-law function of $M_{o}$. 
In this paper we 
present cumulative frequency-size distributions for three new data sets. 
One of these data sets include landslides triggered exclusively by hydrologic
mechanisms, one represents landslides triggered by a single large
earthquake, and one includes landslides triggered by both
hydrologic and seismic mechanisms. Surprisingly, despite
different triggering mechanisms, the frequency-size distributions   
are very similar, with a power-law distribution for large areas and a
flattening
out of the distribution for small areas.

There is also observational evidence that soil moisture exhibits scale-invariant
behavior.
Rodriguez-Iturbe et al. (1995)
have analyzed the frequency-size distribution of patches of soil moisture 
greater than a threshold value and have shown them to have a power-law
cumulative frequency-size distribution. This is analagous to that for landslides.
This suggests that soil moisture patches may be related to landslide
failure areas. 

In Section 2 of this paper we perform spectral analyses of soil moisture
to show that soil moisture is a scale-invariant function. We show that
the classic equation for soil moisture with evapotranspiration proportional
to soil moisture and diffusive dispersion predicts the observed scale-invariance.
In Section 3 we present frequency-size distributions of landslides and argue
that a universal distribution exists, independent of triggering mechanism.
Section 4 attempts to explain the observed frequency-size distribution
of landslides with a slope stability analysis with soil moisture and
topography modeled as scale-invariant functions.

\section{Power spectral analysis of soil moisture data}
We have obtained microwave remotely-sensed soil moisture data collected
from June 10 to June 18, 1992 
at the Washita experimental watershed (Jackson, 1993). 
The watershed received heavy rainfall preceeding the experiment but did not receive
rainfall during the experiment. The data are gridded 
estimates of soil moisture in the top five centimeters of soil
calculated using the algorthm of Jackson and Le Vine (1996). 
Each pixel respresents an area of 200 m x 200 m and the
total area considered is 45.6 km
by 18.6 km. The soil moisture values do not correlate with relief in  
the watershed, which is very small. 
We have computed the one-dimensional power spectrum of each row
of pixels for each image and averaged the power spectrum for each row at
equal frequency values. The power spectrum $S$ is the square of the
Fourier coefficients at each wave number of a Fourier series representation
of the data.
The power spectrum was estimated with the routine
``spctrm'' of Press et al. (1992). The spectra for June 10, 14, and 18 are 
presented from top to bottom in Figure 1. Although the soil moisture
decreased significantly during the course of the experiment, the spectrum has
a nearly constant form. The spectrum is given approximately by a power law
$S(k)\propto k^{-1.8}$, also plotted in Figure 1. The power-law form with
exponent $-1.8$ is chosen to compare with the model presented below. 
A power-law dependence of the power spectral density $S$ on wave number $k$,
$S(k)\propto k^{-2}$,
directly implies scale-invariant behavior. If $\beta=2$ the behavior is
a Brownian walk, with $\beta=1.8$ it is a fractional Brownian walk. In both
cases the behavior is that of a self-affine fractal (Turcotte, 1992).
Although the spectrum appears to flatten out at high
wave numbers, soil moisture variability at the Washita watershed with finer
spatial resolution has been quantified by Rodriguez-Iturbe et al. (1995) 
using similar techniques. They find that the soil moisture variability continues
to be scale invariant
down to the 30 m scale for the more finely sampled data.

In the top image of
Figure 2 we present a color image of the soil moisture at Washita on June 17,
1992. White spaces indicate areas where the watershed is interrupted by
roads or lakes. 
Below it is a synthetic two-dimensional image 
constructed using the
Fourier-filtering method of Turcotte (1992). The mean and variance of the
synthetic moisture field have been chosen to match that of the observed image
and we have taken $\beta=1.8$.
The synthetic image reproduces the correlated structure 
of the real soil moisture image.

Entekhabi and Rodriguez-Iturbe (1994) have proposed a partial differential
equation for the dynamics of the soil moisture field $s(\vec{x},t)$:
\begin{equation}
\frac{\partial s(\vec{x},t)}{\partial t}=D\nabla ^{2}s(\vec{x},t)-
\eta s(\vec{x},t) +\xi(\vec{x},t)
\end{equation}
In this equation, the evapotranspiration rate $\eta$ is assumed
to be constant in space and time. Soil moisture disperses in the soil 
according to a diffusion equation, and rainfall input $\xi(\vec{x},t)$
is modeled as a random function
in space and time. Variations in rainfall from place to place
cause spatial variations in soil moisture that are damped by the effects of
diffusion and evapotranspiration. Without spatial variations in rainfall input, 
there are no variations in soil moisture according to this model.
As pointed out by Rodriguez-Iturbe et al. (1995), the
variability at the small scale of the Washita watershed is not likely to
be the result of spatial variations in rainfall. 

An alternative approach to this problem which will generate spatial 
variations in soil moisture at these small scales assumes that the
evapotranspiration rate is a random function in space and time.
This models spatial and temporal variations in evapotranspiration resulting 
from variable atmospheric conditions and heterogeneity in soil, topography,
and vegetation characteristics. The resulting equation is
\begin{equation}
\frac{\partial s(\vec{x},t)}{\partial t}=D\nabla ^{2}s(\vec{x},t)-
\eta(\vec{x},t)s(\vec{x},t)
\end{equation}
This equation is a variant of the Kardar-Parisi-Zhang equation 
(Kardar et al., 1986) which has
has received a great deal of attention in the physics literature.
This equation is nonlinear and can only be solved numerically. 
Amar and Family (1989) have solved this equation and have
found that the solutions are 
scale-invariant and fractal with a Hausdorff measure of
approximately 0.4, independent of the relative strength of the diffusion and
noise terms. This implies that the one-dimensional power spectrum 
has a power-law dependence on wave number with exponent $\beta=1.8$.
The similarity between the field generated by equation (2) and the soil
moisture observed in the Washita images suggests that equation (2) may 
be capturing much of the essential dynamics of soil moisture at these
scales. 

Rodriguez-Iturbe et al. (1995) quantified the scale-invariance of soil
moisture variations by computing the cumulative frequency-size distribution
of patches of soil with moisture levels higher than a prescribed value.
They found that
the cumulative frequency-size distribution had a power-law function on
area with exponent of approximately $-0.8$.
The number of soil patches $N$ depended on the area according to
\begin{equation}
N(>A)\propto A^{-0.8}
\end{equation}
This is a fractal relation. We have determined the equivalent 
distribution for the synthetic soil moisture field illustrated in Figure 2.
The result is plotted in Figure 3. The same distribution that was observed
in real soil moisture fields is obtained. The same distribution has also
been observed for cumulus cloud fields (Pelletier, 1997).  
 
\section{Observed frequency-size distributions of landslides}
Fuyii (1969),
Whitehouse and Griffiths (1983), Ohmori and Hirano (1988), Sugai (1990),
Sasaki (1991), and Hovius et al. (1997) have all presented evidence that
landslide frequency-size distributions are power-law functions of area.
Some have presented cumulative distributions while others have presented
noncumulative distributions. Care should be taken when comparing distributions
to ensure that the same type of distribution is being used. In this section
we present cumulative frequency-size distributions of landslide area 
from three areas. The results suggest that cumulative frequency-size 
distributions are remarkably similar despite different triggering mechanisms.  

\subsection{Landslides in Japan}
A data set of 3,424 landslides with areas larger than 10$^{4}$ m$^{2}$ in
the Akaishi Ranges, central Japan, have been compiled by Ohmori and Sugai (1995). 
Landslide masses were selected for measurement from 1:20,000 scale
aerial photographs and by field survey. On 1:25,000 scale topographic maps
the length and area of each landslide was measured with a digitizer.    
In this region landslides occur as a result of both heavy rainfall and
strong seismicity. The distribution of landslides is not uniform over the
study area but correlates well with bedrock lithology. For this reason, the
authors have divided their data set into 13 subsets each with a similar lithology.
We have computed the cumulative
frequency-size distribution of landslides in each of
the lithologic units that contained at least 100 landslides. The data set
is believed to be complete for landslides with areas greater than
10$^{-2}$ km$^{2}$.
In Figure 4 cumulative frequency-size distributions, the number of landslides with
an area greater than $A$, are plotted as a function of $A$ 
for seven lithologic units. Areas for these landslides include the
runout zone. The distributions plotted in Figure 4 are remarkably similar
between the different lithologic units. For large landslides, the distribution
is well characterized by a power law with an exponent of approximately $-2$,
that is
\begin{equation}
N(>A)\propto A^{-2}
\end{equation}
This power-law trend for large landslides was chosen based on visual 
correlation with the landslide data. Thus, the exponent $-2$ is only 
approximate. 
The observed frequency-area distribution flattens out for areas less than
about  
10$^{-1}$ km$^{2}$, thus  
few landslides occur with small areas.
Note that this change in the distribution occurs at an area that is
an order of magnitude larger
than the resolution of the catalog. Therefore, it does not appear to be an 
artifact of completeness of the data set.

It can be argued that 
a better estimate of the initial failure area is provided by the
square of the width of the landslide since the length of the landslide
includes the runout distance. 
The data set from Japan includes both estimates of total area as well as
the approximate length and width of each landslide. This allows us the
opportunity to compare the distribution of total landslide area to the
distribution of the square of the width. In Figure 5 the cumulative 
distributions of the total area and the width squared are plotted for
one of the lithologic units of the Japan data set. 
The upper line is the cumulative statistics for total area and the lower
line is the cumulative statistics for width squared.
This figure suggests that the total landslide area scales in
approximately the same way as the initial failure area. The two distributions
were compared for the other lithologic units
of the Japan data set and results similar to those of Figure 5 were obtained.

\subsection{Landslides in Bolivia}
Two sets of landslide areas were mapped in adjacent watersheds in the 
Yungas region
of the Eastern Cordillera, Bolivia between Lake Titicaca and the city of Guanay. 
The main river channels flowing through each of the watersheds are the Challana 
and an unnamed tributary of the Challana.  Because very few people inhabit this 
region, most of the landslides (more than 95\%) were not anthropogenically 
influenced. The watersheds are located in a relatively aseismic region of the 
Andes; therefore, all of the landslides are thought to have been hydrologically 
triggered during the wet season which extends from November through April. 
Based on multitemporal aerial photographic coverage and limited 
dendrochronology field studies, the landslide data sets represent an interval of 
time of about 20-35 years. Scanned aerial photographs at approximately 1:50,000 
scale were used for mapping.  Each photographic image was georegistered to a 
Thematic Mapper image using a Universal Transverse 
Mercator projection.  Landslide
mapping was only undertaken in suitable portions of the watershed below 
treeline. This excluded cloud covered, shadowed, and agricultural areas.  The 
full extent of each landslide was mapped including the runout zone.  Since 
individual trees can be discerned in the mapped areas, the resolution 
of the photographs is about 3 to 10 meters.

The cumulative frequency-size distributions of landslides in these data sets
are presented in Figure 6. Distributions similar to those obtained with the
data from Japan were observed. Large landslides in the two areas match power-law
distributions from equation (4) 
with exponents $-1.6$ and $-2$, determined by visual correlation. 
A flattening of the distribution
at small areas was also observed. As with the Japan data set, this rolloff is
at a scale much larger than the resolution of the data set. 

\subsection{Landslides triggered by the 1994 Northridge, California earthquake}
The January 17, 1994 Northridge, California earthquake triggered more than 
11,000 landslides over an area of 10,000 km$^{2}$. Harp and Jibson (1995)
mapped landslides triggered by the earthquake. They used 
1:60,000-scale aerial photography 
taken the morning of the earthquake by the U.S. Air Force. The 
subsequently digitized photos were supplemented by field work. It was
esimated that the inventory is nearly complete down to landslides of about 5 m
on a side below which a significant number may have been missed. The distribution
is plotted in Figure 7. Large landslides have a power-law dependence on area 
with an exponent of approximately $-1.6$. Again, there is a rolloff 
for small areas. As with the Bolivian and Japanese data sets, 
this rolloff
is at a scale much larger than the resolution scale of the data set and is
not an artifact of an incomplete catalog.   

\section{Modeling of landslide failure with realistic topography and soil moisture}
Landslides occur when the shear 
stress exceeds a threshold value given approximately
by the Coulomb failure criterion
(Terzaghi, 1962)
\begin{equation}
\tau_{f}=\tau_{0}+(\sigma-u)\tan\phi
\end{equation}
where $\tau_{0}$ is the cohesive strength of the soil, $\sigma$ is the 
normal stress on the slip plane, $u$ is the pore pressure, and $\phi$
is the angle of internal friction.  
Landslides are initiated in places where $\tau_{f}$ is greater than a threshold
value. The movement of the soil at the point of instability increases the
shear stress in adjacent points on the hillslope causing failure of a
connected domain with shear stress larger than the threshold value. 
To model the 
landslide instability and, in particular, the frequency-size distribution
of landslides, it is therefore necessary to
model the spatial variations of $\tau_{0}$, $\sigma$, $u$, and $\phi$.
The variables 
$\tau_{0}$ and
$u$ are primarily dependent on soil moisture for a homogeneous
lithology and a slip plane of constant depth. 
The dependence of each of these variables on soil moisture 
has been approximated using power-law functions
(Johnson, 1984). The shear stress and normal stress are 
linearly proportional to soil moisture through the
weight of water in the soil. The shear stress and normal stress are also
trigonometric functions of the local slope. 
Based on the results given in Section 2 the soil moisture will be modeled
as a two-dimensional fractional Brownian walk 
with $\beta=1.8$. Topography will 
be modeled as a Brownian walk.
Power spectral analyses of one-dimensional
transects of topography have shown topography to be a scale-invariant function
analagous to soil moisture. The power spectrum has a power-law dependence
with an exponent $\beta=2$ in a variety of tectonic and erosional
environments (Mandelbrot, 1975; Sayles and Thomas, 1978;
Turcotte, 1987; Huang and Turcotte, 1989). Thus it is a Brownian walk.
This scale-invariant behavior
is applicable only over a certain range of scales. At scales smaller
than the support area of a drainage network, for instance, the surface is
not dissected by channels and the hillslope is smooth. To model the small-scale
smoothness of topography as well as the scale-invariant behavior at larger
scales, we have constructed a two-dimensional surface with one-dimensional
power spectrum $S(k)\propto k^{-2}$ with the Fourier-filtering method of
Turcotte (1992). At small scales, the synthetic topography has been planarly
interpolated. A shaded-relief example of the model topography is illustrated
in Figure 8 along with its contour plot. The plot has 128 x 128 grid points
with interpolation below a scale of 8 pixels. The result of
the interpolation is clearly
identified as piecewise linear segments
in the transects along the boundaries of the plot. The slope
corresponding to this model of topography is illustrated in Figure 9.
Below a scale of 8 pixels, the slopes are constant. Above this scale, the
slopes are a two-dimensional Gaussian white noise. 
This follows from the
fact that our model for topography at large scales, a Brownian 
walk with $S(k)\propto k^{-2}$, 
can be defined as the summation of a Gaussian
white noise time series. White noise means that
adjacent values are totally uncorrelated. The contour map of the slope
function shows that there are no contours smaller than 8 x 8 pixels. 

The shear stress necessary for failure is a complex function of soil moisture
and slope. However, to show how landslide areas may be associated with areas of 
simultaneously high levels of soil moisture and steep slopes, we will 
assume a threshold shear stress criterion proportional to the 
product of the soil moisture and the slope. In addition, we will assume
that slope and soil moisture are uncorrelated. 
A grid of synthetic soil moisture and
topography of 512 x 512 grid points was constructed according to the models
described above. The domains where the product of the soil moisture and the
topography were above a threshold value are shown in Figure 10. The threshold
value was chosen such that only a small fraction of the region was above
the threshold. Figure 11 shows the
cumulative frequency-size distribution of the regions
above threshold, our model landslides. It can be seen that at large 
areas a power-law distribution with an exponent of $-1.6$,
similar to that observed for landslides, is
observed. Distributions obtained with different realizations of the synthetic 
random fields had exponents of $1.6 \pm 0.1$
fit to the landslides above $A=10$. 
For values of the threshold that resulted in only a small fraction
of the lattice being above threshold, 
the form of the distribution was independent of the value
of the threshold. 
The exponent of this distribution is more negative than that of
the soil moisture patches of Figure 3. This results from the less correlated
slope field ``breaking up'' some of the large soil moisture patches so
that there are fewer large landslides relative to small ones than there
are large soil moisture patches relative to small ones. The effect of the
smooth topography at small scales acts to flatten the frequency-size
distribution below the power-law trend. This means that 
the slope function is completely correlated at these scales.
The slope tends to fail as a unit more often than it would if the
slope was not constant on these scales. This results in fewer small 
avalanches and a flattening out of the distribution for small areas.    
The flattening out of the observed landslide distributions
at small areas can also be associated with 
a minimum thickness necessary for failure at depth.   

The effect of strong ground motion from earthquakes is to 
to lower the shear stress necessary for failure (Newmark, 1965). 
This does not alter
the frequency-size distribution of landslides according to our model
since the form of the distribution is independent of the value of the
threshold. This is consistent with the observation of Section 3 that
the cumulative frequency-size distribution is independent of triggering
mechanism. 

\section{Discussion}
There has been a great deal of interest in the physics community in 
systems exhibiting ``avalanches.''
Much of this interest stems from the theory
of self-organized criticality recently introduced by
Bak et al. (1988). These authors proposed the idea that
many systems are driven to a state in which energy is dissipated in events
which have a power-law distribution of sizes. The type example of such a
system is a growing sandpile. Sand grains are dropped on a pile continuously
and are lost from the edge of the pile in discrete sand slides. A few sand slides
are large and many are small. Experimental studies of real sand piles suggest
that the frequency-size distribution of these slides may in fact be power-law
(Held et al., 1990; Bretz et al., 1992; Rosendahl et al., 1994; Densmore et al.,
1997). Rothman et al. (1993)
has presented evidence for power-law statistics in the cumulative frequency-size
distribution of thicknesses of turbidites. It should also be noted that they
observed a flattening of the distribution at small thicknesses similar to
that which we have observed for small areas
in the frequency-size distribution of landslides.
The results of Rothman et al. (1993) have been interpreted in terms of
the theory of self-organized criticality. Self-organized criticality
has also been proposed as model for earthquakes (Bak and Tang, 1989) and
landslides (Noever, 1993). The model we
have presented appears to be an alternative hypothesis for the dynamics
of landslides. 
The behavior of
our model is a result of soil moisture dynamics that are independent
of topography. In the sandpile model
self-organization of the sandpile into a critical state results from the 
continuous input of energy into the system and the action
of large and small sand slides. In that model 
there is no additional field that affects the   
system behavior. In contrast, in our model soil moisture
acts as a additional field
that affects the system dynamics. 

Since soil moisture is known to be an important factor in the landslide 
instability and large landslides obey power-law statistics, it is likely that
patches of soil moisture above a threshold value, 
which also obey power-law statistics, can be associated
with landslides through the model we have discussed. 
This observation suggests that continuous remote-sensing
of soil moisture, together with a digital elevation model, are necessary
for successful landslide hazard assessment. A continuous 
monitoring program of landslide-prone areas could also provide 
valuable data for the better understanding of landslide failure. For example,
in the model we presented, landslide failure was chosen to be the product
of slope and soil moisture. As noted, the correct failure criterion is
likely to be a complex function of the slope and the soil moisture. 
This function could be determined empirically using 
historical records of landslides if high resolution soil moisture data and 
topography are available for the time immediately preceding
each landslide. 
 
\acknowledgements
We wish to thank Peggy O'Neil for providing us with the soil moisture data
and Toshihiko Sugai for providing us with the landslide data from Japan. 

Amar, J.G., and Family, F., 1989. Numerical solution of a continuum equation
for interface growth in 2+1 dimensions. Phys. Rev. A, 41: 3399-3402.

Bak, P., Tang, C., and Wiesenfeld, K., 1988. Self-organized criticality.
Phys. Rev. A, 38: 364-374.

Bak, P., and Tang, C., 1989. Earthquakes as a self-organized critical phenomenon.
J. Geophys. Res., 94: 15,635-15,637.

Beck, A.C., 1968. Gravity faulting as a mechanism of topographic adjustment.
NZ J. Geol. Geophys., 11: 191-199.

Bretz, M., Cunningham, J., Kurczynski, P., and Nori, F., 1992. Imaging of
avalanches in granular materials. Phys. Rev. Lett., 69: 2431-2434.  

Brooks, S.M., and Richards, K.S., 1994. The significance of rainstorm variations
to shallow translational hillslope failure. Earth Surf. Proc. Landforms, 19:
85-94.

Caine, N., 1980. The rainfall-intensity-duration control of shallow landslides
and debris flows. Geografiska Annaler, 62: 23-27.

Committee in Ground Failure Hazards, 1985. Reducing Losses from Landsliding
in the United States. Commission on Engineering and Technical Systems,
National Research Council, Washington, D.C., 41 pp.

Densmore, A.L., Anderson, R.S., McAdoo, B.G., and Ellis, M.A., 1997. Hillslope
evolution by bedrock landslides. Science, 275: 369-372.

Entekhabi, D., and Rodriguez-Iturbe, I., 1994. Analytical framework for the
characterization of the space-time variability of
soil moisture. Adv. Water Resour., 17: 35-46.

Fukuoka, M., 1980. Landslides associated with rainfall. Geotechnical Engineering,
11: 1-29.

Fuyii, Y., 1969. Frequency distribution of the magnitude of landslides
caused by heavy rainfall. J. Seismol. Soc. Japan, 22: 244-247.

Grove, J.M., 1972. The incidence of landslides, avalanches, and floods in eastern
Norway during the little ice age. Artic and Alpine Res., 4: 131-138.

Harp, E.L., and Jibson, R.L., 1995. 
Inventory of landslides triggered by the 1994 Northridge,
California earthquake, US Geol. Surv. Open File Report 95-213.

Held, G.A., Solina II, D.H., Keane, D.T., Haag, W.J., Horn, P.M., and
Grinstein, G., 1990. Experimental study of critical mass fluctuations
in an evolving sandpile. Phys. Rev. Lett., 65: 1120-1123.   

Hovius, N., Stark, C.P., and Allen, P.A., 1997.
Sediment flux from a mountain belt derived by landslide mapping. Geology

Huang, J., and Turcotte, D.L., 1989.
Fractal mapping of digitized images: Applications
to the topography of Arizona and comparisons with synthetic images.
J. Geophys. Res., 94: 7491-7495.

Innes, J.L., 1983. Lichenometric dating of debris-flow deposits in the Scottish
highlands, Earth Surf. Processes, 8: 579-588.

Jackson, T.J., 1993. Washita '92 Data Sets, high density diskette, version 12/20.

Jackson, T.J., and Le Vine, D., 1996.
Mapping surface soil moisture using an aircraft-based
passive microwave instrument: algorithm and example. J. Hydrol., 184: 85-99.

Jibson, R.W., 1996. Use of landslides for paleoseismic analysis. Engineering
Geology, 43: 291-323.

Jibson, R.W., and Keefer, D.K., 1988. Landslides triggered by earthquakes
in the central Mississippi Valley, Tennessee, and Kentucky. US Geol.
Surv. Professional Paper
1336-C.

Johnson, A.M., 1984. Debris Flow, In: D. Brunsden and D.B. Prior (Editors),
Slope Instability. Wiley, New York, pp. 257-361. 

Johnson, K.A., and Sitar, N., 1990. Hydrologic conditions leading to 
debris-flow initiation. Can. Geotech. J., 27: 789-801.

Kardar, M., 
Parisi, G., and Zhang, Y.-C., 1986.
Dynamic scaling of growing interfaces. Phys. Rev.
Lett.,
56: 889-892.

Keefer, D.K., 1984. Landslides caused by earthquakes. Geol. Soc. Am. Bull., 95:
406-421.

Keefer, D.K., 1994. The importance of earthquake-induced landslides to long-term
slope erosion and slope-failure hazards in seismically active regions.
Geomorphology, 10: 265-284.

Mandelbrot, B., 1975. Stochastic models for the earth's relief, the shape and the
fractal dimension of coastlines, 
and the number-area rule for islands. Proc. Nat. Acad. Sci. U.S.A., 72: 3825-3828.

Newmark, N.M., 1965. Effects of earthquakes on dams and embankments.
Geotechnique, 15: 139-160.

Noever, D.A., 1993. Himalayan sandpiles. Phys. Rev. E, 47: 724-725.

Ohmori, H., and Hirano, M., 1988. Magnitude, frequency, and geomorphological
significance of rocky mud flows, landcreep, and the collapse of steep
slopes. Z. Geomorph. N.F., 67: 55-65.

Ohmori, H., and Sugai, T., 1995. Toward geomorphometric models for estimating
landslide dynamics and forecasting landslide occurrence in Japanese
mountains. Z. Geomorph. N.F., 101: 149-164.

Pelletier, J.D., 1997. Kardar-Parisi-Zhang scaling of the height of the 
convective boundary layer and the fractal structure of cumulus cloud
fields. Phys. Rev. Lett., 78:2672-2675.

Pitts, J., 1983. The temporal and spatial development of landslides in the
Axmouth-Lyme Regis Undercliffs National Nature Reserve, Devon. Earth
Surf. Processes, 8: 589-603.

Press, W.H., Teukolsky, S.A., Vetterling, W.T., and Flannery, B.P., 1992.
Numerical Recipes in C: The Art of Scientific Computing (2nd edn.). New York,
Cambridge
Univ. Press, 994 pp.

Rodriguez-Iturbe, I., Vogel, G.K., Rigon, R., Entekhabi, D., and Rinaldo, A., 1995.
On the spatial
organization of soil moisture fields. Geophys. Res. Lett., 22: 2757-2760.

Rosendahl, J., Vekic, M., and Kelley, J., 1994. Persistent self-organization of
sandpiles. Phys. Rev. E, 47: 1401-1404.

Rothman, D.H., Grotzinger, J., and Flemings, P., 1993. Scaling in turbidite
deposition, J. Sed. Petrol. A, 64: 59-67.

Sasaki, Y., 1991. Fractals of slope failure number-size distribution.
J. Japan. Soc. Engineering Geol., 32: 1-11.

Sayles, R.S., and Thomas, T.R., 1978. Surface topography as a non-stationary
random process, Nature, 271: 431-434.

Seed, H.B., 1968. Landslides during earthquakes due to soil liquefaction.
Am. Soc. Civil Engineers. J. Soil Mech. Found. Div. 94: 1055-1122.

Sugai, T., Ohmori, H., and Hirano, M., 1994. Rock control on magnitude-frequency distributions of landslides.
Trans. Japan. Geomorph. Union, 15: 233-251.

Terzaghi, K., 1962.
Stability of steep slopes on hard unweathered rock. Geotechnique,
12: 251-270.

Tibaldi, A., Ferrari, L., and Pasquare, G., 1995. Landslides triggered by 
earthquakes and their relations with faults and mountain slope geometry:
an example from Ecuador. Geomorphology, 11: 215-226.

Turcotte, D.L., 1987. 
A fractal interpretation of topography and geoid spectra on the
Earth,
Moon, Venus, and Mars. J. Geophys. Res., 92: 597-601.

Turcotte, D.L., 1992. Fractals and chaos in geology and geophysics. 
New York, Cambridge Univ. Press, 221 pp. 

Whitehouse, I.E., and Griffiths, G.A., 1983.
Frequency and hazard of large rock
avalanches in the central Southern Alps, New Zealand, Geology, 11: 331-334.

Wieczoreck, G.F., 1987. Effect of rainfall intensity and duration on debris
flows
in central Santa Cruz Mountains, California. In:
J.E. Costa and G.F. Wieczorek (Editors), Debris Flows/Avalanches:
Process, Recognition, and Mitigation, Geological Society of America.
Rev. Eng. Geol., 7: 93-104.

Wieczoreck, G.F., 1996. 
Landslide triggering mechanisms, In: A.K. Turner and R.L. Schuster (Editors),
Landslides: Investigation and
Mitigation. Transportation Research Board, National Academy of Sciences, 
Washington, D.C., Special Report 247, 
pp. 76-90.
\newpage
\section*{Figure Captions}

Figure 1: Average one-dimensional power spectrum of soil moisture on
(top) June 10, (middle) June 14, and (bottom) June 18, 1992 in the
Washita experimental watershed. The spectra are offset so that they
can be plotted on the same graph.

Figure 2: (a) Color map of microwave remotely-sensed estimates of
soil moisture on June 17, 1992 at the Washita experimental watershed.
(b) Color map of a synthetic soil moisture field with the power
spectrum $S(k)\propto k^{-1.8}$ with the same mean and variance as
the remotely-sensed data.  

Figure 3: Cumulative frequency-size distribution of patches of
soil moisture larger than a threshold value for the synthetic soil
moisture field with power spectrum $S(k)\propto k^{-1.8}$. 

Figure 4: Cumulative frequency-size distribution of landslides in 
6 lithologic zones in Japan. The distributions are well characterized
by $N(>A)\propto A^{-2}$ above aproximately 0.1 km$^{2}$.

Figure 5: The upper curve is the cumulative frequency-size distribution
for one of the lithologic zones in Japan as illustrated in Figure 4.
The lower curve is the data for the same set of landslides with the area
defined to be the width of the landslide squared. 

Figure 6: Cumulative frequency-size distribution of landslides in
two areas in Bolivia.

Figure 7: Cumulative frequency-size distribution of landslides triggered
by the 1994 Northridge, California earthquake.   

Figure 8: Synthetic model of topography used in the model of Section 4.
The topography is a scale-invariant function with $S(k)\propto k^{-2}$
above a scale of eight lattice sites and is planarly interpolated below
that scale. 

Figure 9: Slope model corresponding to the topographic model of Figure 8.
The slopes are white noise above a scale of 8 lattice points and constant
below that scale.

Figure 10: Contour map of the product of the synthetic soil moisture 
field with the synthetic slope function. Areas inside the contour loops 
respresent model landslides.  

Figure 11: Cumulative frequency-size distribution of model landslides.
The distribution compares favorably to the distributions of real landslides. 
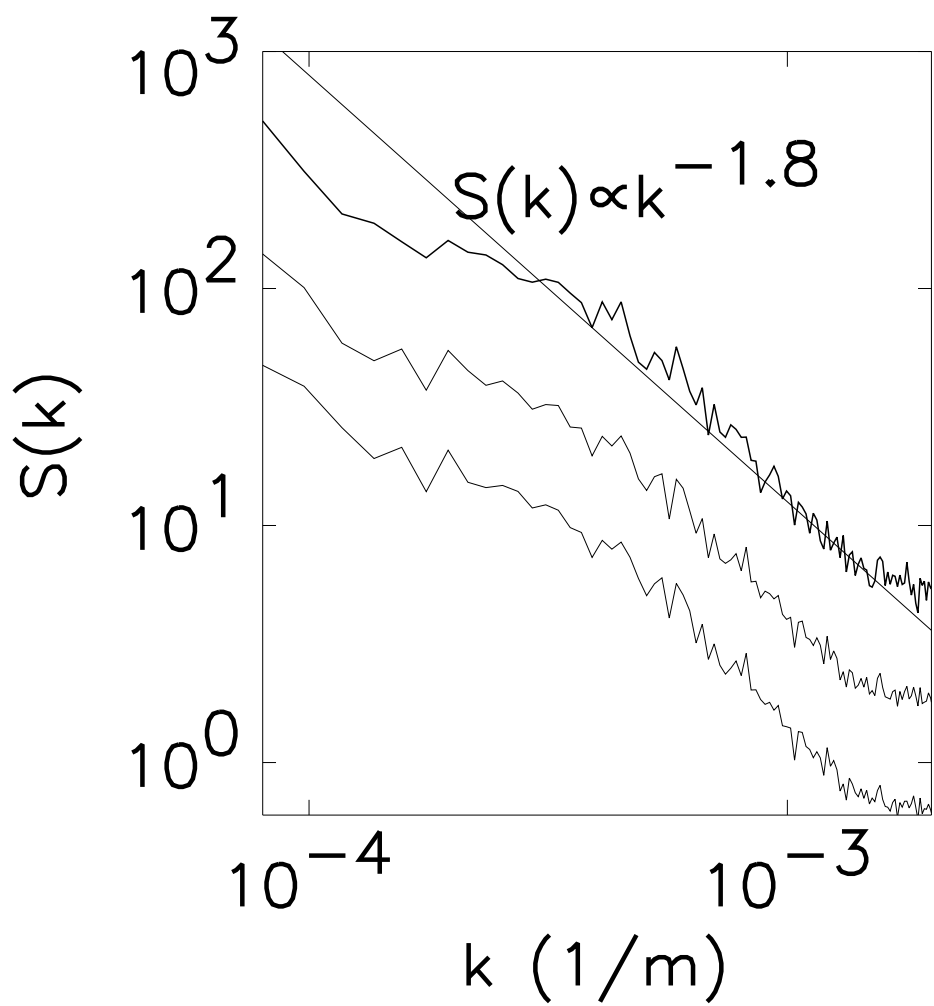
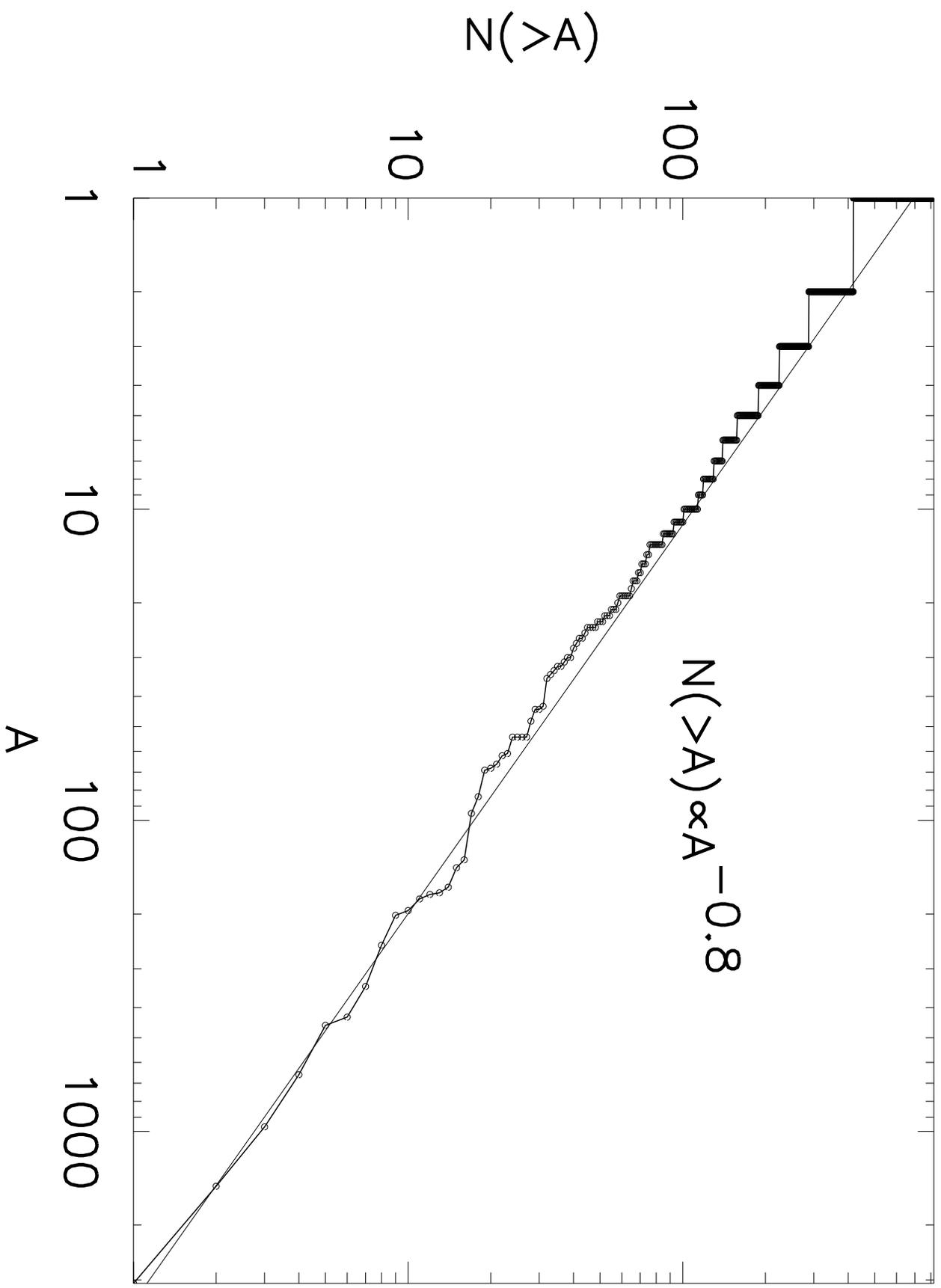
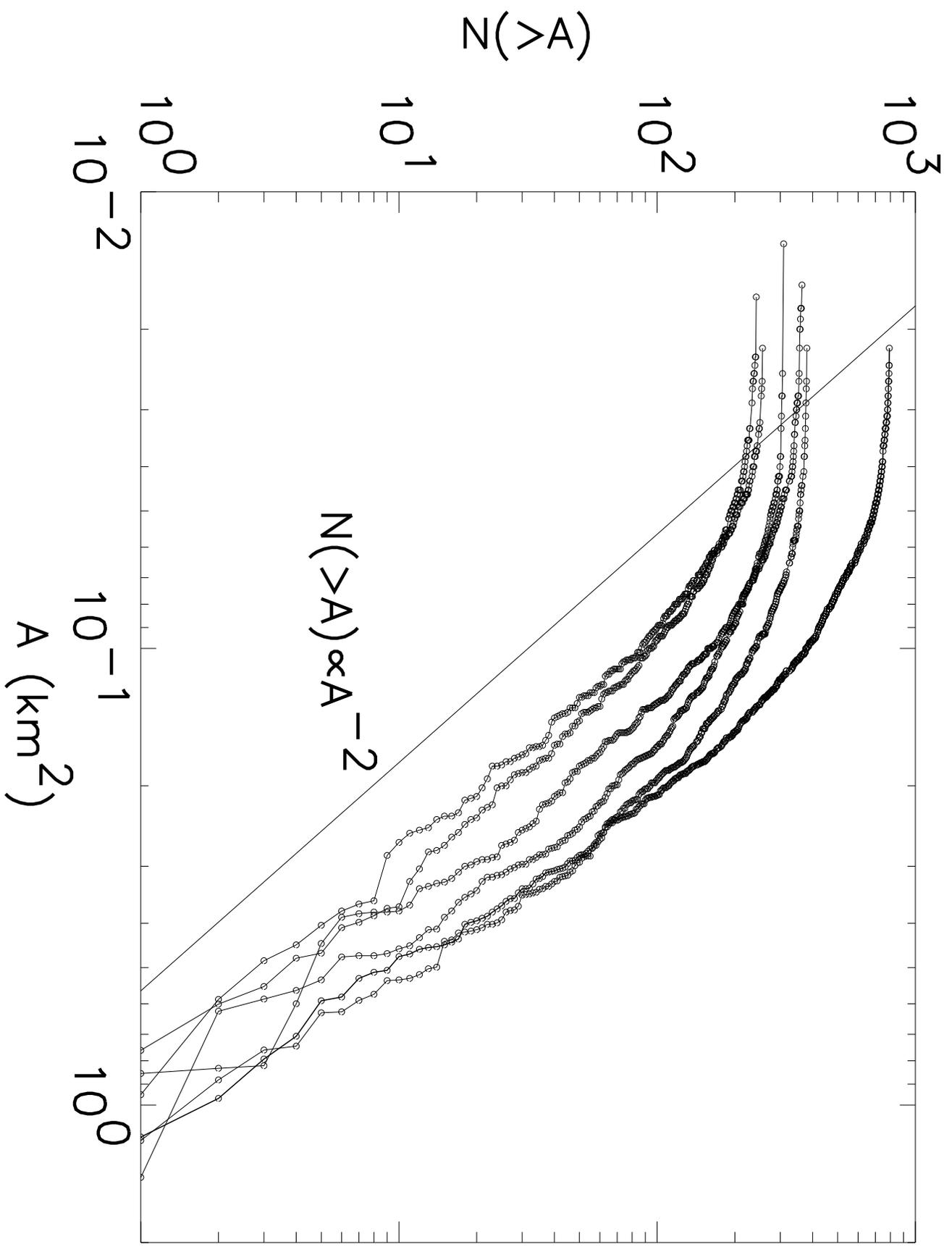
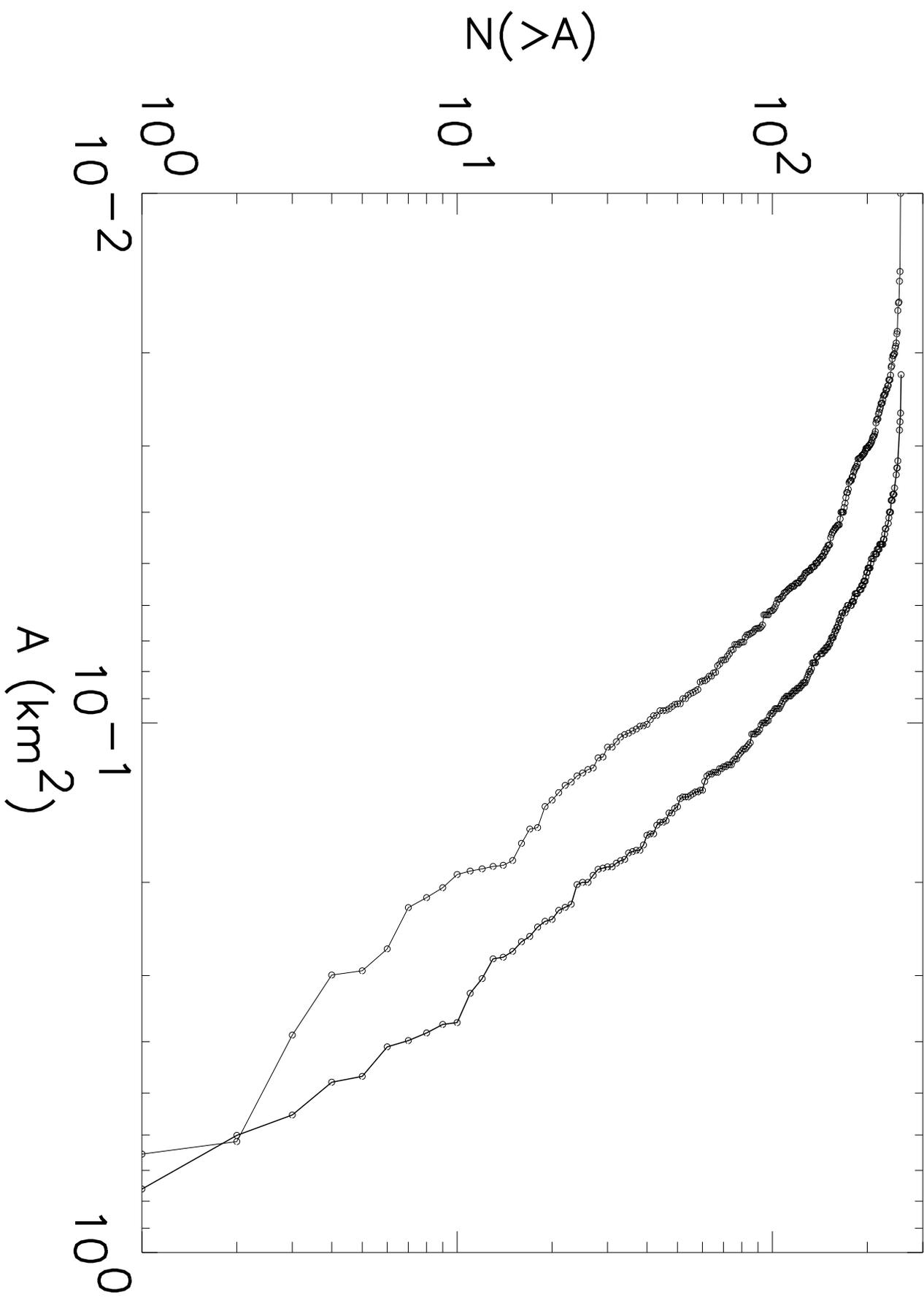
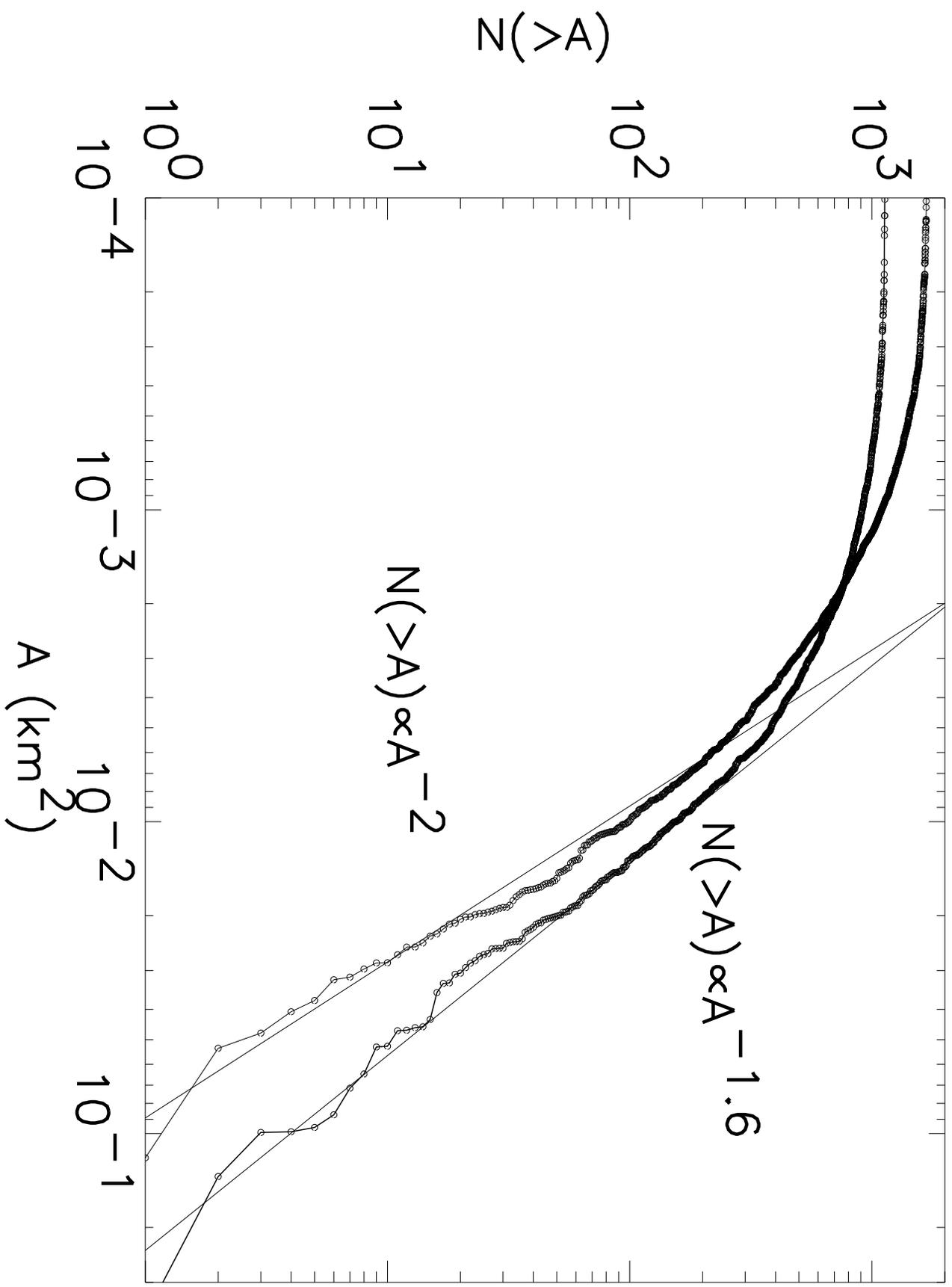
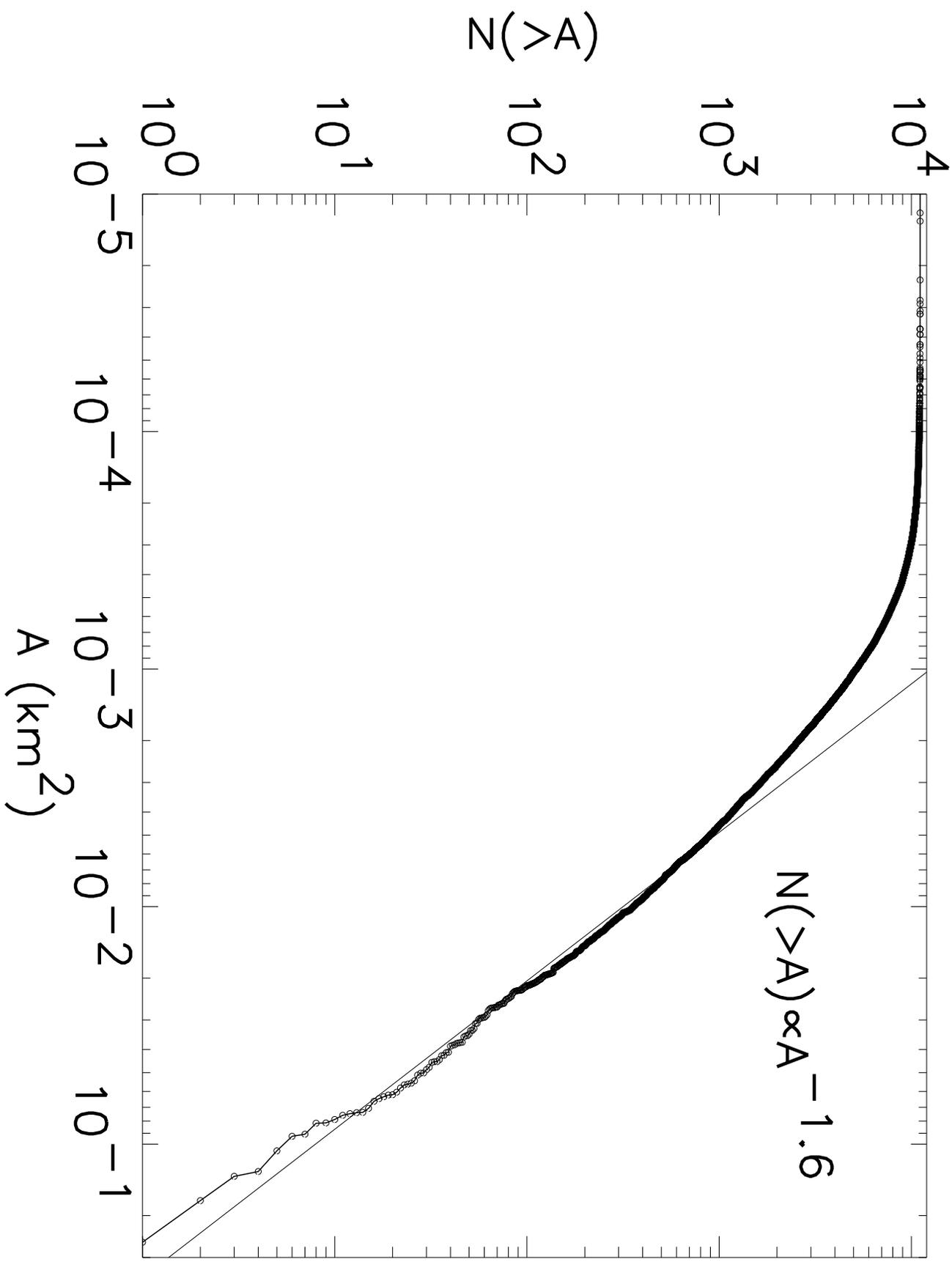
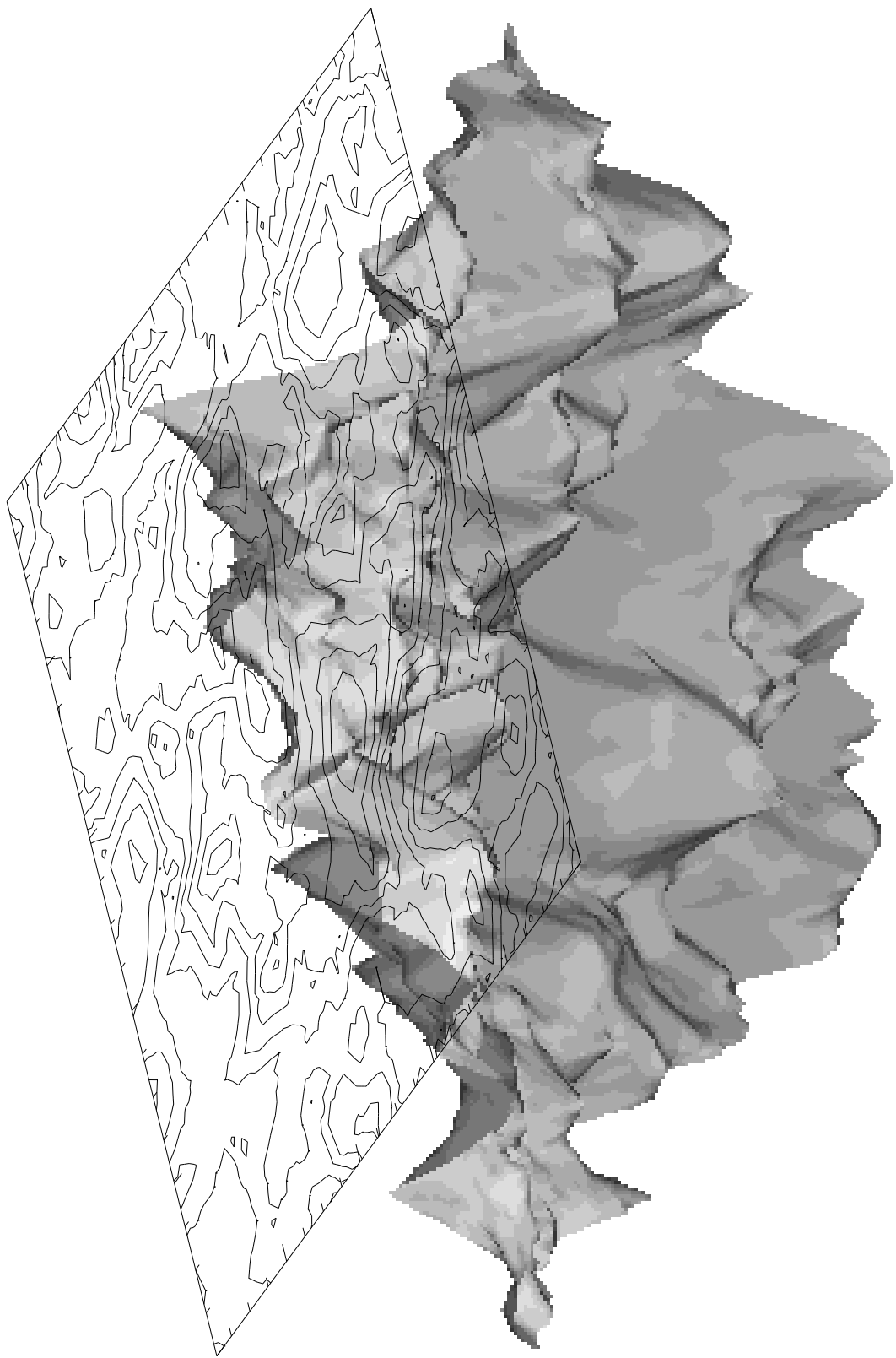
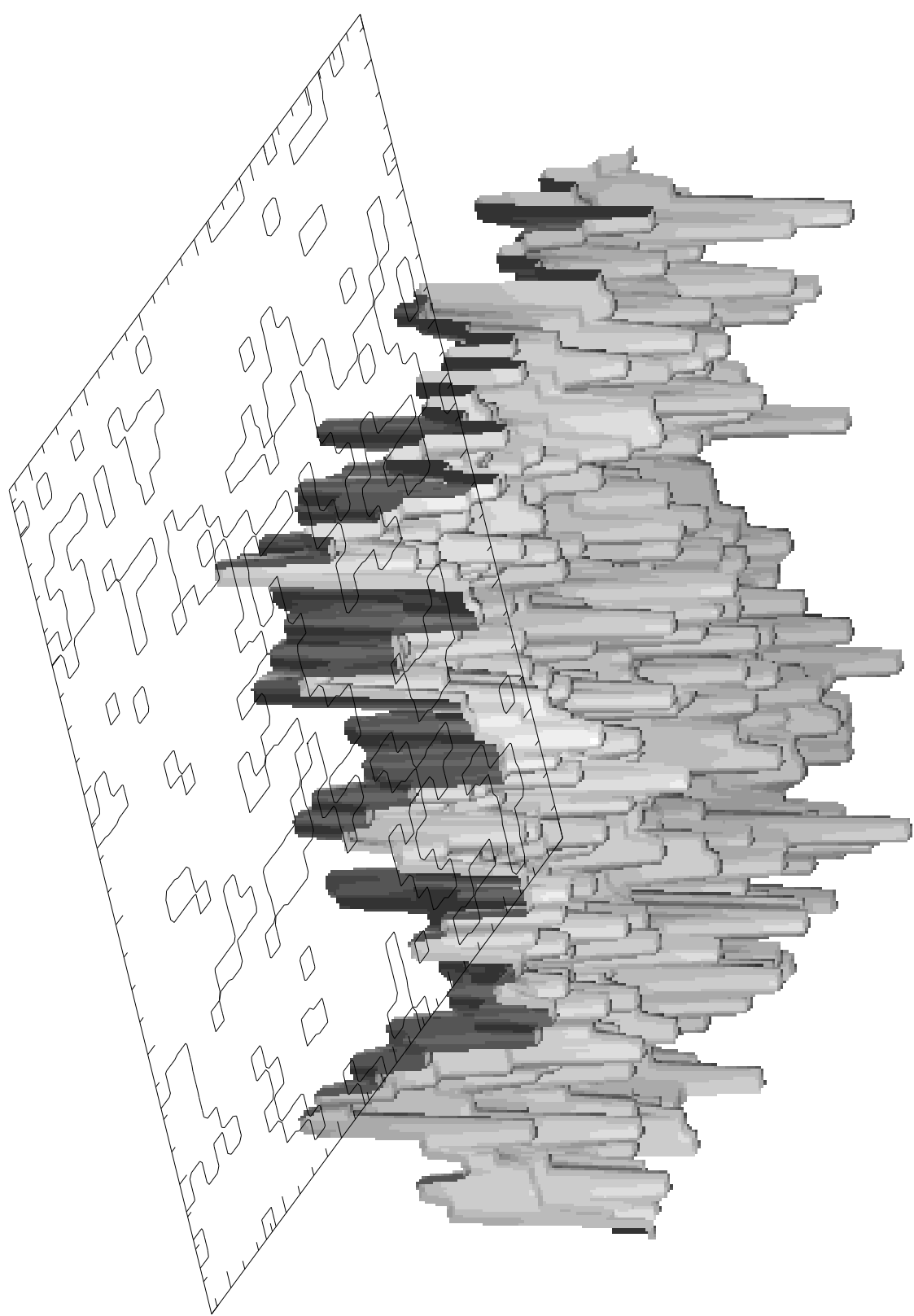
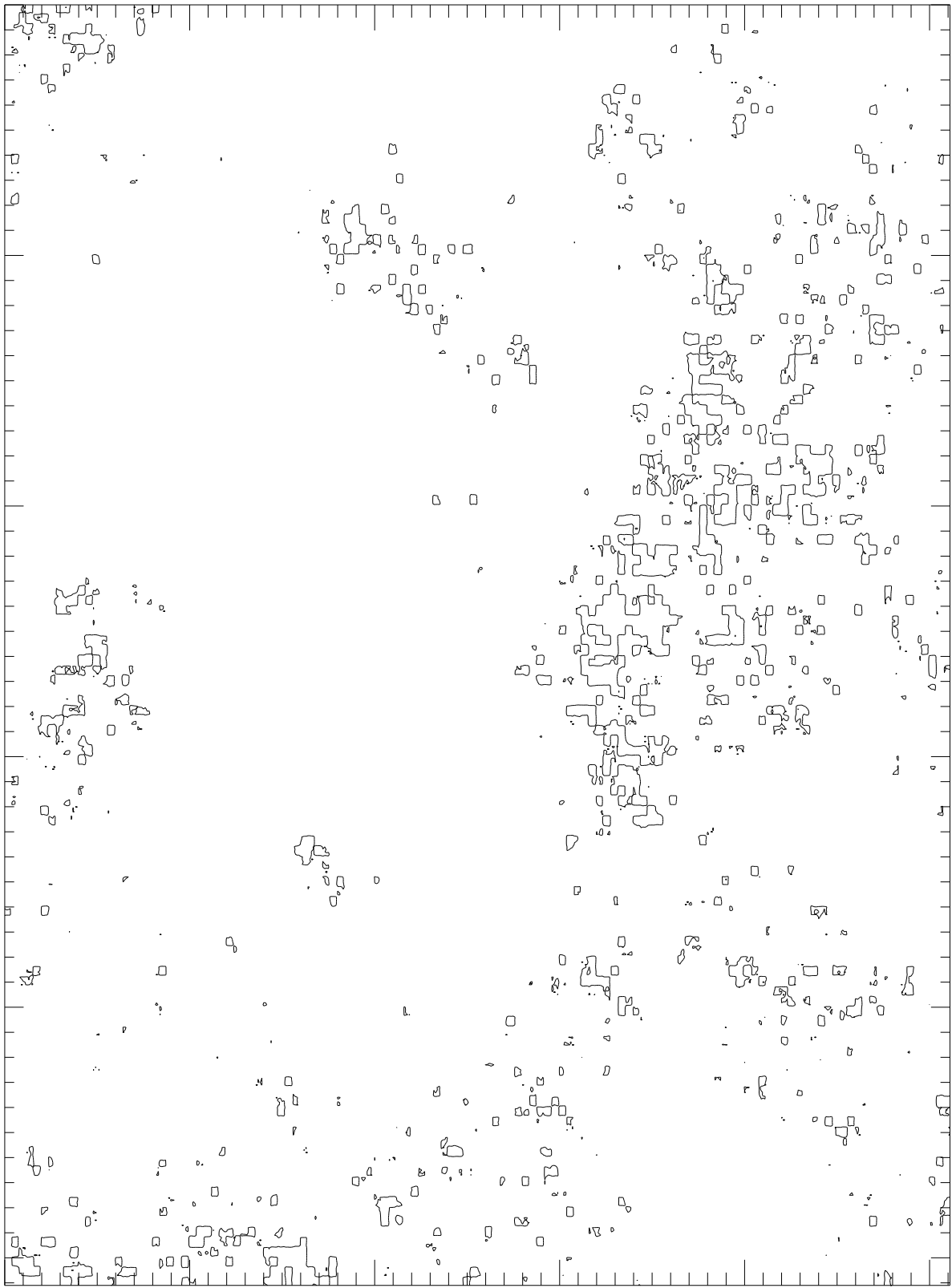
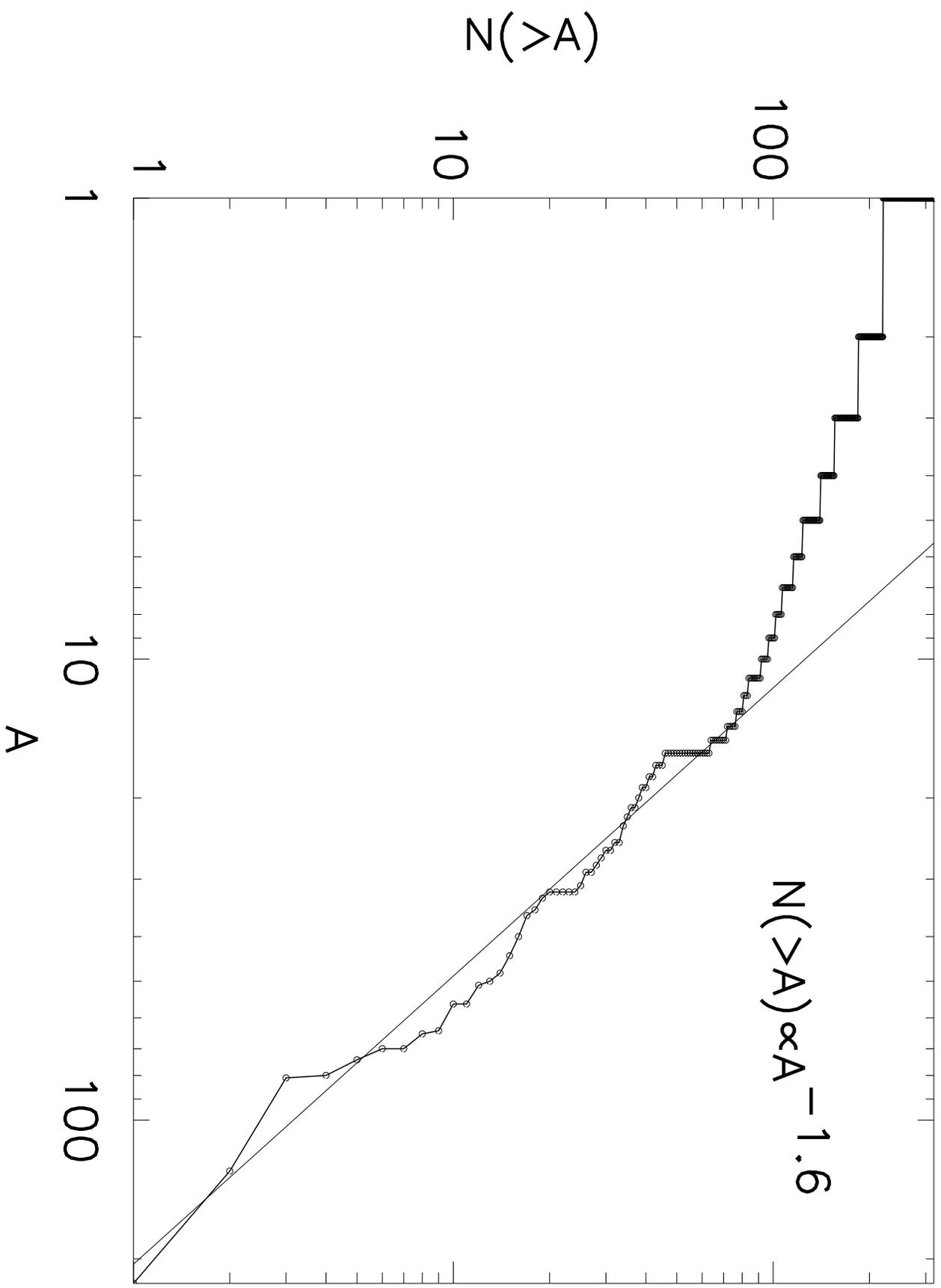
\end{document}